\documentclass[prd,preprint,tightenlines,floatfix,showpacs,preprintnumbers,
nofootinbib,eqsecnum,superscriptaddress]{revtex4}

 \usepackage[dvips,final]{graphicx}
  \usepackage{amssymb}
   \usepackage{amsmath}
    \usepackage{amsfonts}
     \usepackage{epsfig}
      \usepackage{bm}

\def\Pom{{\bf I\!P}}
\newcommand{\bsigma}{\mbox{\boldmath $\sigma$}}

\newcommand{\bp}{\mbox{\boldmath $p$}}
\newcommand{\bq}{\mbox{\boldmath $q$}}
\newcommand{\br}{\mbox{\boldmath $r$}}

\newcommand{\bk}{\mbox{\boldmath $k$}}
\newcommand{\bb}{\mbox{\boldmath $b$}}
\newcommand{\be}{\mbox{\boldmath $e$}}
\newcommand{\bn}{\mbox{\boldmath $n$}}
\newcommand{\bM}{\mbox{\boldmath $M$}}
\newcommand{\ket}[1]{| {#1} \rangle}
\newcommand{\bra}[1]{\langle {#1} |}

\newcommand{\half}{{1\over 2}}

\begin{document}

\thispagestyle{empty} \preprint{\hbox{}} \vspace*{-10mm}

\title{Exclusive photoproduction of $J/\psi$ \\ in proton-proton
and proton-antiproton scattering}

\author{W. Sch\"afer}
\email{Wolfgang.Schafer@ifj.edu.pl}
\affiliation{Institute of Nuclear Physics PAN, PL-31-342 Cracow,
Poland} 
\author{A. Szczurek}
\email{Antoni.Szczurek@ifj.edu.pl}
\affiliation{Institute of Nuclear Physics PAN, PL-31-342 Cracow,
Poland} 
\affiliation{University of Rzesz\'ow, PL-35-959 Rzesz\'ow,
Poland}

\date{\today}

\begin{abstract}
Protons and antiprotons at collider energies are a source
of high energy Weizs\"acker--Williams photons. 
This  may open a possibility to study exclusive photoproduction of
heavy vector mesons at energies much larger than possible at the HERA
accelerator. Here we present
a detailed investigation of the exclusive 
$J/\psi$ photoproduction in proton-proton (RHIC, LHC) and
proton-antiproton (Tevatron) collisions. We calculate several 
differential distributions in $t_1, t_2, y, \phi$, as well as
transverse momentum distributions of $J/\Psi$'s.
We discuss correlations in the azimuthal angle between outgoing protons
or proton and antiproton as well as in the ($t_1, t_2$) space.
Differently from electroproduction experiments, here both
colliding beam particles can be a source of photons, and
we find large interference terms in azimuthal angle distributions
in a broad range of rapidities of the produced meson.
We also include the spin--flip parts in the electromagnetic vertices.
We discuss the effect of absorptive corrections on various 
distributions.
Interestingly, absorption corrections induce a charge asymmetry
in rapidity distributions, and are larger for $p p$ reactions 
than for the $p \bar p$ case.
The reaction considered here constitutes an important nonreduceable
background in recently proposed searches for odderon exchange.
\vspace{1cm}
\begin{center}
{\it{dedicated to Kolya Nikolaev on the occasion of his 60th birthday}}
\end{center}
\vspace{1cm}
\end{abstract}

\pacs{13.87.Ce, 13.60.Le, 13.85.Lg}

\maketitle

\section{Introduction}

The diffractive photoproduction of $J/\psi$--mesons has been recently a
subject of thorough studies at HERA \cite{ZEUS_JPsi,H1_JPsi}, 
and serves to elucidate
the physics of the QCD pomeron and/or the small--$x$ gluon 
density in the proton (for a recent review and references, see
\cite{INS06}). Being charged particles, protons/antiprotons 
available at e.g. RHIC, Tevatron and LHC are a source of high energy
Weizs\"acker--Williams photons, and photoproduction processes are
also accessible in hadronic collisions.
Hadronic exclusive production mechanisms 
of mesons at central rapidities in $pp$ collisions
were intensively studied in the 1990-ies at energies of a few tens of GeV 
\cite{Central_experiment}, and raised much theoretical interest
for their potential of investigating exotic hadronic states 
(see e.g. \cite{Central_theory}).
Recently there was interest in describing diffractive exclusive
production of heavy scalar \cite{KMRS04,PST07_chic}
and pseudoscalar \cite{SPT06_etap} mesons in terms of off-diagonal 
unintegrated gluon distributions, which may provide 
insight into the related diffractive production mechanism of the Higgs 
boson (\cite{KMR97,Saclay} and references therein).
A purely hadronic mechanism for the exclusive production 
of $J/\Psi$ mesons in proton-proton and 
proton-antiproton collisions was suggested as a candidate
in searches for yet another exotic object of QCD, the elusive
odderon exchange \cite{SMN91,BMSC07}. 
In order to identify the odderon exchange one has to consider all
other possible processes leading to the same final channel
which in the context of the searches for the odderon will constitute 
the unwanted background.
One of such processes (and perhaps the only one at the level of fully
exclusive $J/\Psi$-production) is pomeron-photon or photon-pomeron fusion
\cite{KMR_photon,KN04,BMSC07}, which we study in this 
communication at a more detailed level than available in
the literature.
We feel that its role as a background for odderon
searches warrants a more detailed analysis including energy
dependence and differential distributions of the photoproduction
mechanism in hadronic collisions. As will be discussed, 
the process considered here is interesting also in its own rights.

An important concern of our work are absorption effects.
More often than not absorption effects are either completely ignored
or included as a multiplicative reduction factor, which is simply
wrong for many observables (like distributions in $t_1$ or $t_2$) 
as we shall show in our paper.  
We think this point requires broader public spread 
as it often appears to be forgotten or ignored.
We present a detailed analysis of several differential
distributions in order to identify the absorption effects.
We also put a special emphasis on interference phenomena. 
We will discuss also more subtle phenomena 
like the spin flip in the electromagnetic vertices and 
a charge asymmetry by comparing differential distributions in 
proton-proton and proton-antiproton exclusive $J/\psi$ production.

In this work, we do not include a possible odderon contribution.
We wish to stress, that the photoproduction mechanism of
exclusive $J/\Psi$'s must exist without doubt, and does not die
out as energy increases. 
A related purely hadronic (Odderon) contribution with the same properties 
has not been unambiguously identified in other experiments, and hence 
cannot be estimated in a model-independent way.
While certain QCD-inspired toy-models for CP-odd multigluon t-channel 
exchanges exist, they do not allow reliable calculations of 
hadronic amplitudes.
In practice the magnitude of the corresponding Born amplitude 
strongly depend on the details of how to treat gluons
in the nonperturbative domain, as well as on the modeling of
proton structure.  Furthermore, the energy dependence of the full 
(beyond the Born approximation, and beyond perturbation theory) 
amplitude is unknown and it can even not be
excluded that this contribution would vanish with rising energy.
It is not the issue of our paper to further discuss such models, 
we rather think that in the search for an odderon one should take 
further initiative, if substantial deviations from the more conservative 
physics discussed here are found.

\section{Amplitudes and cross sections}

\subsection{$2 \to 3$ amplitude}
\label{2_to_3}

\begin{figure}[!h]    %
\includegraphics[width=\textwidth]{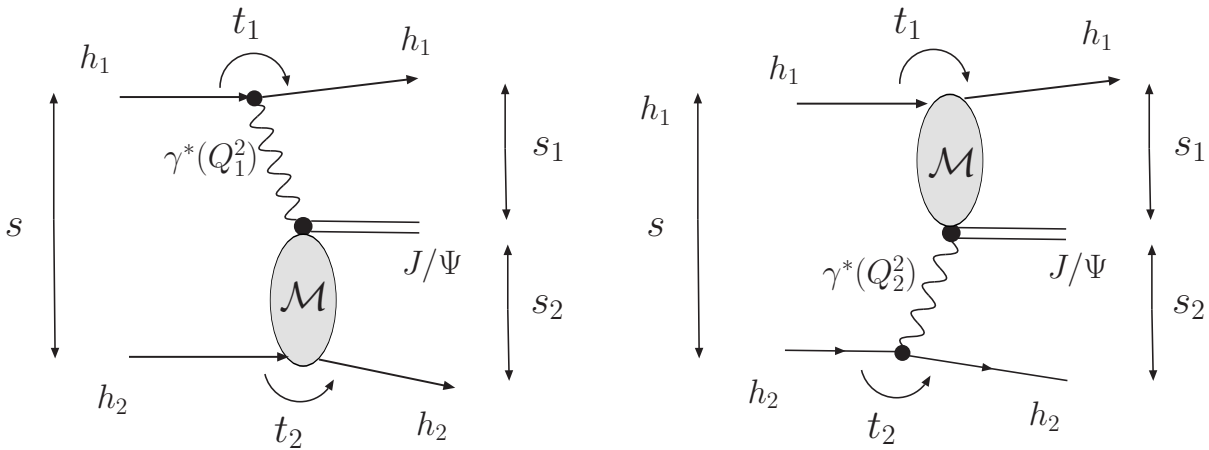}
   \caption{\label{fig:diagram_photon_pomeron}
   \small  The sketch of the two mechanisms considered in the present paper:
$\gamma \Pom$ (left) and $\Pom\gamma$ (right).
Some kinematical variables are shown in addition.}
\end{figure}

Here we present the necessary formalism for the calculation
of amplitudes and cross--sections. The basic mechanisms 
are shown in Fig.\ref{fig:diagram_photon_pomeron}. 


The distinctive feature, when compared to photoproduction
in lepton--hadron collisions, is that now both
participating hadrons can serve as the source of the photon,
and it is necessary to take account of the interference
between the two amplitudes. 
Due to the Coulomb singularities in the photon exchange parts of
the amplitude, the electromagnetic vertices involve only very small,
predominantly transverse momentum transfers. Their effect 
is fully quantified by the well--known electromagnetic
Dirac-- and Pauli form factors of the nucleon.
Regarding the photoproduction amplitude, we try to be as far 
as possible model independent, and take advantage of the
precise knowledge of diffractive vector meson production over
a broad energy range available from experiments at HERA.
The amplitude for the $2 \to 3$ process of 
Fig.\ref{fig:diagram_photon_pomeron} can be decomposed as 
\begin{eqnarray}
{\cal M}_{h_1 h_2 \to h_1 h_2 V}^
{\lambda_1 \lambda_2 \to \lambda'_1 \lambda'_2 \lambda_V}(s,s_1,s_2,t_1,t_2) &&= 
{\cal M}_{\gamma \Pom} + {\cal M}_{\Pom \gamma} \nonumber \\
&&= \bra{p_1', \lambda_1'} J_\mu \ket{p_1, \lambda_1} 
\epsilon_{\mu}^*(q_1,\lambda_V) {\sqrt{ 4 \pi \alpha_{em}} \over t_1}
{\cal M}_{\gamma^* h_2 \to V h_2}^{\lambda_{\gamma^*} \lambda_2 \to \lambda_V \lambda_2}
(s_2,t_2,Q_1^2)   \nonumber \\
&& + \bra{p_2', \lambda_2'} J_\mu \ket{p_2, \lambda_2} 
\epsilon_{\mu}^*(q_2,\lambda_V)  {\sqrt{ 4 \pi \alpha_{em}} \over t_2}
{\cal M}_{\gamma^* h_1 \to V h_1}^{\lambda_{\gamma^*} \lambda_1 \to \lambda_V \lambda_1}
(s_1,t_1,Q_2^2)  \, . \nonumber \\
\label{Two_to_Three}
\end{eqnarray}
The outgoing protons lose only tiny fractions $z_1,z_2 \ll 1$
of their longitudinal momenta.In terms of their transverse momenta
$\bp_{1,2}$ the relevant four--momentum transfers squared are
$t_i = - (\bp_i^2 + z_i^2 m_p^2)/(1-z_i) \, , i = 1,2$,
and $s_1 \approx (1 -z_2) s$ and $s_2 \approx (1-z_1) s$ are the
familiar Mandelstam variables for the appropriate subsystems. 
Due to the smallness of the photon virtualities, denoted by $Q_i^2 = -t_i$, 
\footnote{Of course here the notation $Q_i^2 = t_i$ applies only to 
the photon lines.} 
it is justified to neglect the contribution from
longitudinal photons, recall that $\sigma_L(\gamma^* p \to J/\Psi p)
/\sigma_T(\gamma^* p \to J/\Psi p) \propto Q^2/m_{J/\Psi}^2$
\cite{H1_JPsi,INS06}.
Then, the amplitude for emission of a photon of 
transverse polarization $\lambda_V$, and transverse momentum
$\bq_1 = - \bp_1$,  entering eq.(\ref{Two_to_Three}) reads:
\begin{eqnarray} 
\bra{p_1', \lambda_1'} J_\mu \ket{p_1, \lambda_1} 
\epsilon_{\mu}^*(q_1,\lambda_V)
&&= { (\be^{*(\lambda_V)} \bq_1)  \over \sqrt{1-z_1}} 
\, {2 \over z_1} \, \chi^\dagger_{\lambda'} 
\Big\{  F_1(Q_1^2) 
- {i \kappa_p F_2(Q_1^2) \over 2 m_p}  
( \bsigma_1 \cdot [\bq_1,\bn]) \Big\} \chi_\lambda \, ,
\nonumber \\
\end{eqnarray}
here $\be^{(\lambda)} = -(\lambda \be_x + i \be_y)/\sqrt{2}$,
$\bn || \be_z$ denotes the collision axis,
and $\bsigma_1/2$ is the spin operator for nucleon $1$, $\chi_\lambda$
is its spinor. $F_1$ and $F_2$ are the Dirac-- and
Pauli electromagnetic form factors, respectively.
Here we have given only that part of the current, which gives rise
to the logarithmic $dz/z$ longitudinal momentum spectrum of photons,
which dominates in the high--energy kinematics considered here.
It is worthwhile to recall, that for a massive fermion, that
includes a spin--flip contribution originating from its
anomalous magnetic moment, $\kappa_p = 1.79$. 
Notice its suppression at small transverse momenta.
The parametrization of the photoproduction amplitude
which we used in practical calculations can be found
in the Appendix. Above we already used the assumption
of $s$--channel--helicity conservation in the 
$\gamma^* \to J/\Psi$ transition, 
which for heavy vector mesons is indeed 
well justified by experiment \footnote{While a trend towards
s-channel helicity violating effects may be visible in the H1 data
\cite{H1_JPsi}, they are surely negligible for our purpose, and
within error bars, consistent with \cite{ZEUS_JPsi} and s-channel
helicity conservation.}
\cite{ZEUS_JPsi,H1_JPsi,INS06}.
In summary we present the $2 \to 3$ amplitude in the form
of a 2--dimensional vector as
\begin{eqnarray} 
\bM(\bp_1,\bp_2) &&= e_1 {2 \over z_1} {\bp_1 \over t_1} 
{\cal{F}}_{\lambda_1' \lambda_1}(\bp_1,t_1)
{\cal {M}}_{\gamma^* h_2 \to V h_2}(s_2,t_2,Q_1^2)   
\nonumber \\
&&
+ e_2 {2 \over z_2} {\bp_2 \over t_2} {\cal{F}}_{\lambda_2' \lambda_2}(\bp_2,t_2)
{\cal {M}}_{\gamma^* h_1 \to V h_1}(s_1,t_1,Q_2^2)  \, . 
\end{eqnarray}
The differential cross section of interest is given in terms of $\bM$ as
\begin{equation}
d \sigma = { 1 \over 512 \pi^4 s^2 } | \bM |^2 \, dy dt_1 dt_2
d\phi \, ,
\end{equation}
where $y \approx \log(z_1 \sqrt{s}/m_{J/\Psi})$ is the rapidity of the 
vector meson, and $\phi$ is the angle between $\bp_1$ and $\bp_2$.
Notice that the interference between the two mechanisms $\gamma \Pom$
and $\Pom \gamma$ is proportional to $e_1 e_2 (\bp_1 \cdot \bp_2)$ 
and introduces a charge asymmetry as well as an angular correlation
between the outgoing protons.
Clearly, the interference cancels out after integrating over $\phi$,
and the so integrated distributions will coincide for $pp$ and
$p\bar{p}$ collisions.

\subsection{Absorptive corrections}
\label{Absorption}

\begin{figure}[!h]    %
\includegraphics[width=\textwidth]{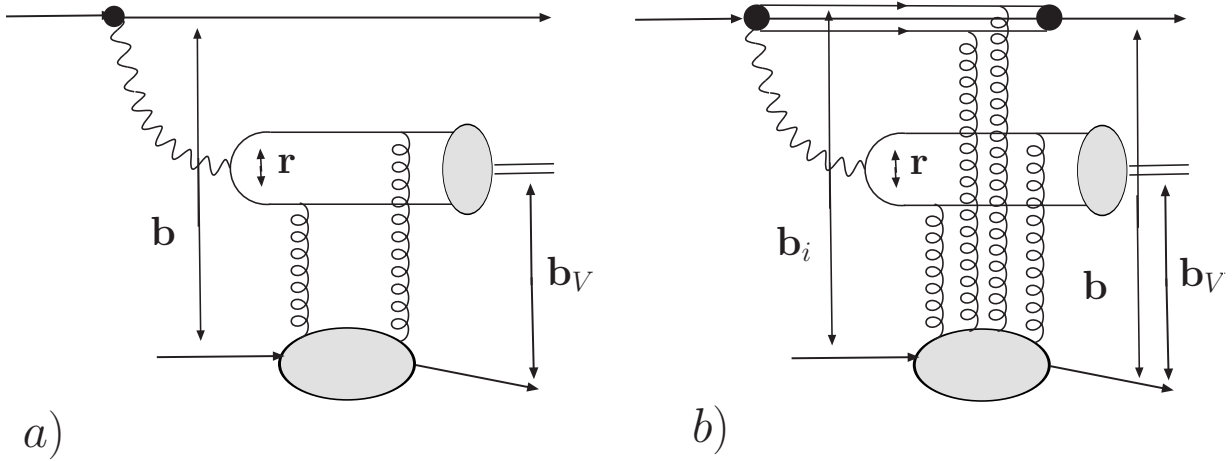}
   \caption{
   \small  Left: the QCD two--gluon exchange mechanism for the Born--level
amplitude. Right: a possible multigluon--exchange contribution that involves 
uncancelled spectator interactions. The impact parameters relevant for 
the discussion are indicated.}
\label{fig:multigluon}
\end{figure}

We still need to correct for a major omission in our description
of the production amplitude. Consider for example a rest frame of
the proton 2 (the target) of the left panel of 
Fig \ref{fig:diagram_photon_pomeron}. Here, the virtual photon may be viewed 
as a parton of proton 1 (the beam), separated from it by a large
distance in impact parameter space. It splits into its $c\bar{c}$--Fock 
component at a large longitudinal distance before the target,
and to obtain the sought for production amplitude we project the 
elastically scattered $c\bar{c}$ system onto the desired $J/\Psi$--final 
state \cite{Kolya_CT}. We entirely neglected
the possibility \cite{Bjorken} that the photon's spectator partons might  
participate in the interaction, and destroy the rapidity gap(s)
in the final state. Stated differently, for the diffractive final state 
of interest, spectator interactions do not cancel, and will affect the
cross section. As a QCD--mechanism consider the interaction
of a $\{c\bar{c}\}_1 \{qqq\}_1$--beam system with the target 
by multiple gluon exchanges (see Fig \ref{fig:multigluon} ). 
Then, for the $J/\Psi$ final state of interest, the interaction of the 
$c \bar{c}$--color singlet state is dominated by small dipole sizes $r_s \sim
4/m_{J /\Psi}$ (the scanning radius of \cite{Kolya_VM}). It can
be exhausted by the minimal two--gluon color--singlet exchange,
and will be quantified by the color--dipole cross
section $\sigma(\br)$ \cite{Kolya_CT,NZ_91}, respectively its non--forward
generalization \cite{NNPZZ,INS_Saturation}.
Let $\bb_V$ be the tranverse separation of the $J/\Psi$
and the target, and $\br$ the size of the $c\bar{c}$--dipole
as shown in Fig \ref{fig:multigluon}. Then, the
$2 \to 3$ amplitude of section \ref{2_to_3} will involve, besides the
vertex for the $p \to \gamma p$ transition, the expectation value
$\bra{J/\Psi} \Gamma^{(0)}(\br,\bb_V) \ket{\gamma}$ of 
\begin{equation}
\Gamma^{(0)}(\br , \bb_V ) = { 1\over 2}  \, \sigma(\br)  \, t_N(\bb_V, B) \, ,
\end{equation}
where  $t_N(\bb,B) =  \exp(-{\bb^2/2B } ) / (2 \pi B ) $ is an optical
density of the target.
Systematic account for the spectator interactions in QCD however is
a difficult problem, as one cannot rely on the
Abramovsky-Gribov-Kancheli (AGK) \cite{AGK} cutting rules, 
when due account for color is taken \cite{NS_AGK}. 
To obtain at least a qualitative account of absorptive corrections 
we restrict ourselves to only a subclass of
absorptive corrections, the 'diffractive cut', which contribution
is model independent \cite{Gribov}. 
Regarding our  $\{c\bar{c}\}_1 \{qqq\}_1$--system, with
$\bb_i$ denoting the constituent quarks' impact parameters, and
$\bb$ the impact parameter of the beam proton, the 
absorbed amplitude in impact parameter space will contain
\begin{eqnarray}
\Gamma(\br , \bb_V,\bb ) &&= { 1\over 2}  \, \sigma(\br)  \, t_N(\bb_V, B)
- {1 \over 4} \sigma(\br) \sigma_{qqq}(\{\bb_i\})  t_N(\bb_V, B)
t_N(\bb,B_{el}) \nonumber \\
&&= \Gamma^{(0)}(\br , \bb_V ) \Big( 1 - \half \sigma_{qqq}(\{\bb_i\})
t_N(\bb, B_{el}) \Big) \to \Gamma^{(0)}(\br , \bb_V ) \cdot S_{el}
(\bb) \, .
\end{eqnarray}
In effect, we  merely multiply the Born--level amplitude $\Gamma^{(0)}$
by the probability amplitude for beam and target to pass through each
other without inelastic interaction.
In momentum space, we obtain the absorbed amplitude as
\begin{eqnarray}
\bM(\bp_1,\bp_2) &&= \int{d^2 \bk \over (2 \pi)^2} \, S_{el}(\bk) \,
\bM^{(0)}(\bp_1 - \bk, \bp_2 + \bk)  \nonumber \\
&&= \bM^{(0)}(\bp_1,\bp_2) - \delta \bM(\bp_1,\bp_2) \, ,
\label{rescattering_term}
\end{eqnarray}
and with 
\begin{equation}
S_{el}(\bk) = (2 \pi)^2 \delta^{(2)}(\bk) - \half T(\bk) \, \, \, ,
\, \, \, T(\bk) = \sigma^{pp}_{tot}(s) \, \exp\Big(-\half B_{el} \bk^2 \Big) \, ,
\end{equation}
the absorptive correction $\delta \bM$ reads 
\footnote{
In the practical calculations below, for Tevatron energies,
we take $\sigma^{p\bar p}_{tot} = 76$ mb, $B_{el} = 17 \, \mathrm{GeV}^{-2}$
\cite{Tevatron}. 
}
\begin{eqnarray}
\delta \bM(\bp_1,\bp_2) = \int {d^2\bk \over 2 (2\pi)^2} \, T(\bk) \,
\bM^{(0)}(\bp_1-\bk,\bp_2+\bk) \, .
\label{absorptive_corr}
\end{eqnarray}
A number of improvements on this result can be expected to be relevant.
Firstly, a more consistent microscopic treatment of spectator
interactions along the lines of \cite{NS_AGK} would be desirable. 
Experience from hadronic phenomenology \cite{KNP} suggests
that at Tevatron energies, the purely elastic rescattering taken into 
account by eq.(\ref{absorptive_corr}) are insufficient, and inelastic
screening corrections will to a crude estimate lead to an enhancement
of absorptive corrections by a factor 
$\lambda \sim (\sigma_{el} + \sigma_{D})/\sigma_{el}$ 
\cite{TerM}. Here $\sigma_D = 2 \sigma_{SD} + \sigma_{DD}$, and
$\sigma_{SD} = \sigma(pp \to pX) \, , \, \sigma_{DD} = \sigma(pp \to XY)$
are the cross sections for single--, and double--diffractive processes,
respectively.
Secondly, also the $\gamma p \to J/\Psi p$ production amplitude will
be affected by unitarity corrections. For example, with increasing
rapidity gap $\Delta y$ between $J/\Psi$ and the target, one
should account for additional $s$--channel gluons, and for sufficiently
dense multiparton systems, the two-gluon exchange approximation for the
$\gamma \to J/\Psi$ transition used above, ultimately becomes inadequate.
For relevant discussions of unitarity/saturation--effects in
diffractive $J/\Psi$--production, see \cite{Saturation}, the
scaling properties of vector meson production in the presence
of a large saturation scale are found in \cite{INS_Saturation}.    
In our present approach, where the production amplitude is taken
essentially from experiment, one must content oneself with the
fact, that (some) saturation effects are effectively contained in our
parametrization, and any extrapolation beyond the energy domain
covered by data must be taken with great caution.

\begin{figure}[!h]    %
\includegraphics[width=\textwidth]{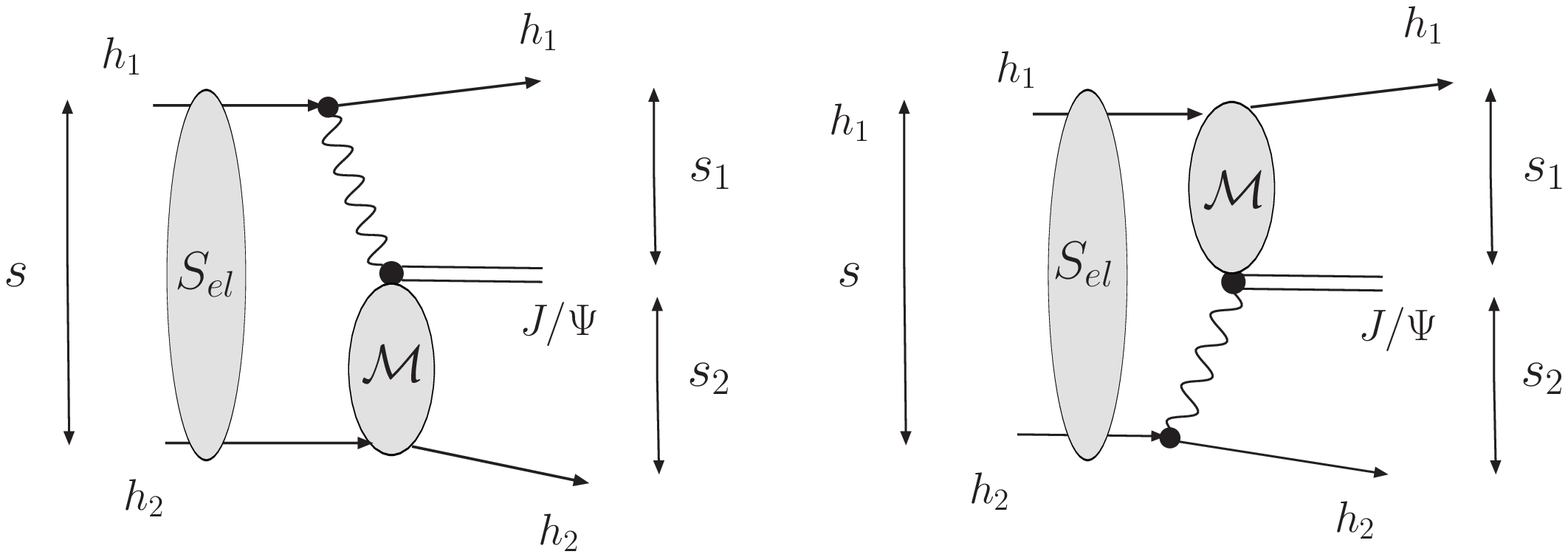}
   \caption{\label{fig:diagram_rescattering}
   \small  A sketch of the elastic rescattering amplitudes effectively
taken into account by eq.(\ref{rescattering_term}).
}
\end{figure}
%

\section{Results}

In this section we shall present results of differential cross sections
for $J/\Psi$ production.
We shall concentrate on the Tevatron energy $W=1960 \textrm{GeV}$,
where such a measurement might be possible even at present.
While in this paper we concentrate on the fully exclusive process
$pp \to pp J/\Psi$, $p\bar p \to p \bar p J/\Psi$, it is important to realize,
that from an experimental point of view there are
additional contributions related to the exclusive production of $\chi_c$
mesons and their subsequent radiative decays to $J/\Psi \gamma$. 
It may be difficult to measure/resolve the soft decay photons and therefore 
experimentally this contribution may be seen as exclusive 
production of $J/\psi$. We note in this context that besides the
scalar $\chi_c(0^{++})$ meson, which exclusive production has been 
discussed in the literature (e.g. \cite{KMRS04} and references
therein), the axial--vector and tensor states
$\chi_c(1^{++})$ and $\chi_c(2^{++})$ have
larger branching fractions into the relevant $J/\Psi \gamma$ channel. 
Although their exclusive production cross sections can be expected to 
be suppressed at low transverse momenta \cite{KMRS04,Yuan}, 
a more detailed numerical analysis is not existing in the 
literature and would clearly go beyond the scope of the present paper.

\subsection{Distributions of $J/\psi$}


\begin{figure}[!h]  
 \centerline{\includegraphics[width=0.5\textwidth]{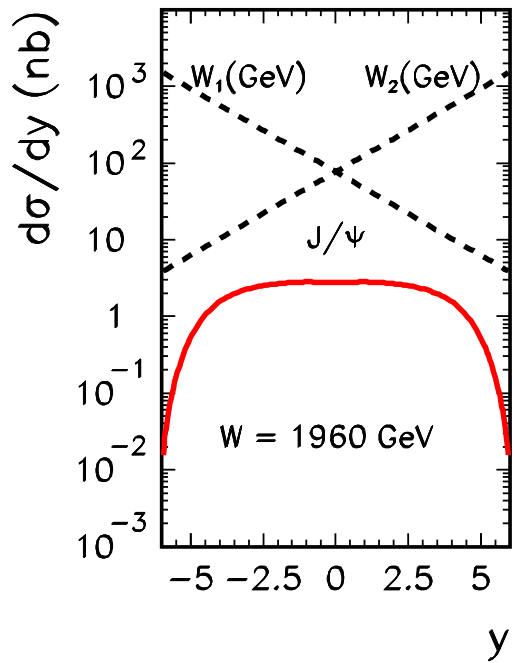}}
   \caption{\label{fig:dsig_dy}
   \small $d \sigma / dy$ as a function of the $J/\Psi$ rapidity ($y$)
 for W = 1960 GeV. For a better understanding of the results we also show (dashed lines)
the subsystem energies $W_{1V}=\sqrt{s_1}$ and $W_{2V}=\sqrt{s_2}$ in GeV.
}
\end{figure}
\begin{figure}[!h]   
 \centerline{\includegraphics[width=0.5\textwidth]{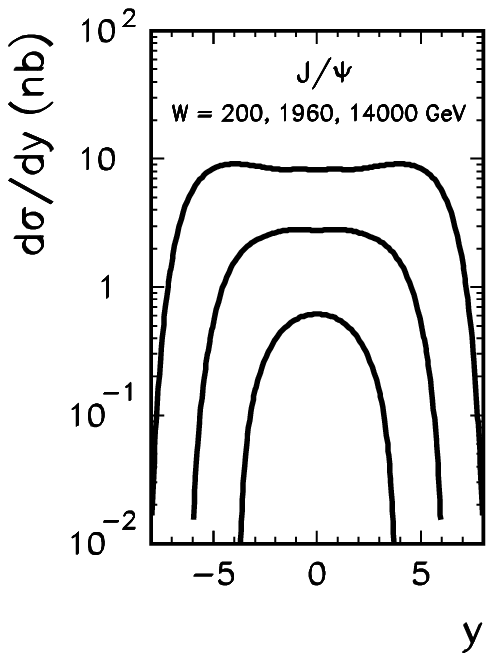}}
   \caption{ \label{fig:dsig_dy_energy}
\small  $d \sigma / dy$ for exclusive $J/\psi$ production
as a function of $y$ for RHIC, Tevatron and LHC energies.
No absorption corrections were included here.}
\end{figure}

Let us start from the rapidity distribution of the $J/\Psi$ shown in
Fig.\ref{fig:dsig_dy}. In the figure we present also the subsystem energies 
$\sqrt{s_1},\sqrt{s_2}$. At $|y| > 3$ the energies of the $\gamma p \to J/\psi p$
or $\gamma \bar p \to J/\psi \bar p$ subprocesses exceed the energy range
explored at HERA. This may open a possibility to study $J/\Psi$ photoproduction
at Tevatron. This is interesting by itself and requires further
detailed studies. In turn, this means that our estimate of
the cross section far from midrapidity region requires extrapolations
above the measured energy domain.
In Fig.\ref{fig:dsig_dy_energy} we collect rapidity distributions
for different energies relevant for RHIC, Tevatron and LHC. We observe an
occurence of a small dip in the distribution at midrapidities at LHC energy.
The shape of the rapidity distribution at LHC energies however relies
precisely on the above mentioned extrapolation of the parametrization
of HERA--data to higher energies.
Clearly a real experiment at Tevatron and LHC would help to constrain 
cross sections for $\gamma p \to J/\psi p$ process.


\begin{figure}[!h]   
\includegraphics[width=0.4\textwidth]{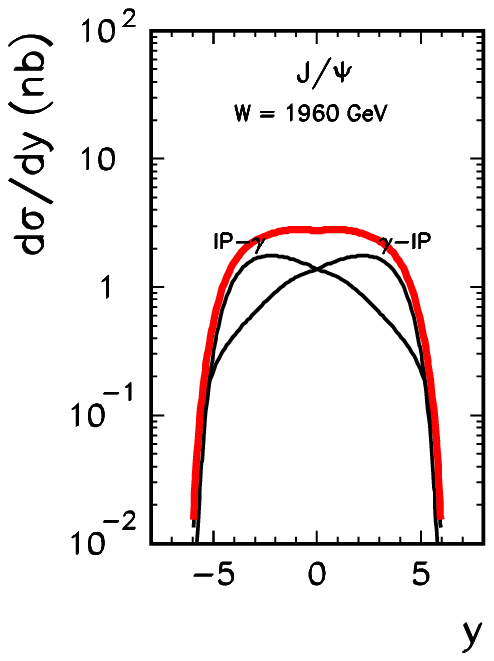}
\includegraphics[width=0.4\textwidth]{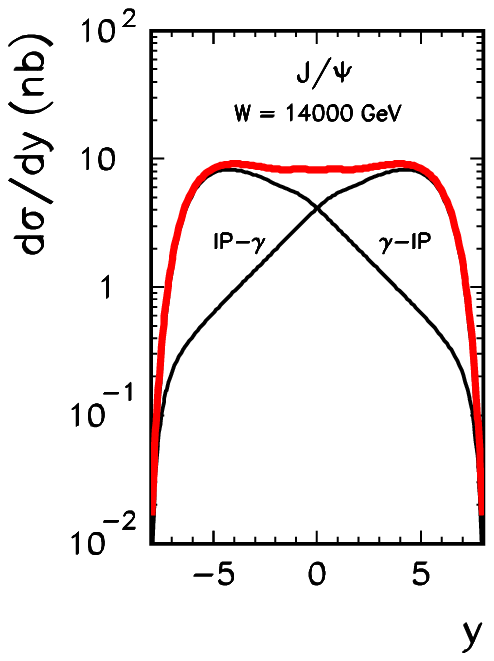}
   \caption{ \label{fig:dsig_dy_deco}
\small  $d \sigma / dy$ as a function of $y$ for the Tevatron and LHC energy.
Individual processes are shown separately. Notice that the beam hadron $h_1$ 
of Fig. \ref{fig:diagram_photon_pomeron} moves at positive rapidities.
No absorption corrections were included here.}
\end{figure}

In order to understand the origin of the small dip at midrapidity at LHC
energy in Fig.\ref{fig:dsig_dy_deco} we show separately the contributions
of the two components ($\gamma \Pom, \Pom \gamma$ exchange)
for Tevatron (left) and LHC (right). We see that at LHC energy the two 
components become better separated in rapidity. This reflects the strong
rise of the $J/\Psi$ photoproduction cross section with energy,
which can be expected to slow down with increasing energy.
Notice that the beam hadron $h_1$ moves along positive rapidities, so
that, for example for the mechanism $\gamma\Pom$ it is the Pomeron
exchange which 'propagates' over the larger distance in rapidity space.
It would be interesting to confront our present simple predictions with
the predictions of the approach which uses unintegrated gluon
distributions -- objects which are/were tested in other high-energy
processes. This will be a subject of our forthcoming studies.


\begin{figure}[!h]   
\includegraphics[width=.5\textwidth]{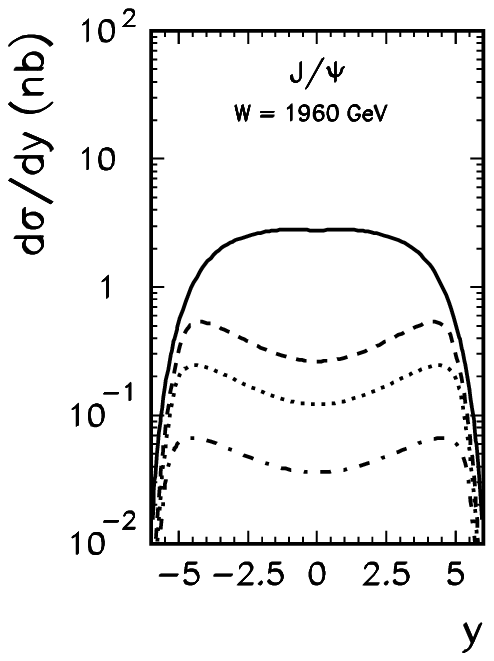}
   \caption{ \label{fig:dsig_dy_tcut}
\small
$d \sigma / dy$ as a function of $y$ for the Tevatron energy and
different upper cuts on $t_1$ and $t_2$: $t_{cut}$ = 0.0 GeV$^2$
(solid),
$t_{cut} = -0.05$ GeV$^2$ (thin solid),
$t_{cut} = -0.1$ GeV$^2$ (dash-dotted),
and $t_{cut} = -0.2$ GeV$^2$ (dashed).}
\end{figure}

\begin{figure}[!hp]     
\includegraphics[width=0.4\textwidth]{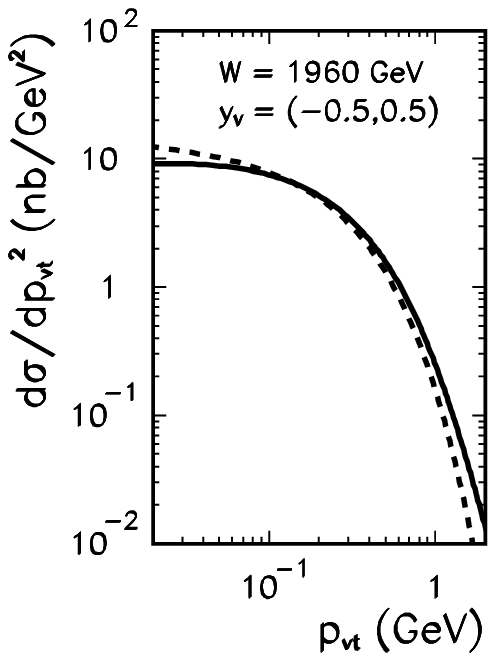}
\includegraphics[width=0.4\textwidth]{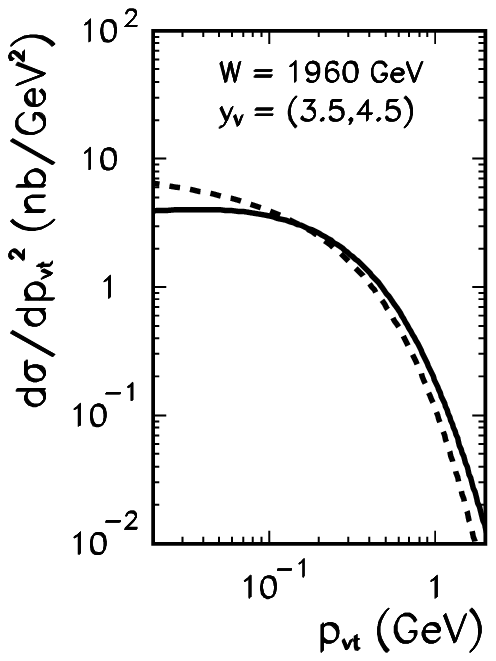}
\caption{ 
The distribution $d\sigma/d \bp_{V}^2$ of $J/\psi$
as a function of $J/\psi$ transverse momentum for different
intervals of rapidity: $-0.5 < y < 0.5$ (left panel) and
$3.5 < y < 4.5$ (right panel) at W = 1960 GeV.
The result for $p \bar p$ collisions is shown by the solid line and the result
for $p p$ collisions by the dashed line.
No absorption corrections were included here.
\label{fig:dsig_dpvt2}}
\end{figure}

Up to now we have not taken into account any restrictions on 
$t_1$ and/or $t_2$.
In practice, it can be necessary to impose upper cuts on the transferred
momenta squared. It is also interesting in the context of searches for odderon,
to see how quickly the cross section for the ``background'' drops with $t_1$
and $t_2$.
In Fig.\ref{fig:dsig_dy_tcut} we show distribution in $J/\psi$ rapidity
for different cuts on $t_1$ and $t_2$.
Clearly, imposing a cut on $t_1$ and $t_2$ removes the 
photon-pole contribution dominant at small momentum transfers.
Even relatively small cut lowers the cross section considerably,
and the dropping of the cross section is much faster than
for the pomeron-odderon exchanges \cite{BMSC07}.
Imposing upper cuts on $t_1$ and $t_2$ will therefore help considerably
to obtain a possible ``odderon-enriched'' sample.
We wish to repeat here that without absorption effects the rapidity
distribution of $J/\psi$ in proton-proton and proton-antiproton
collisions are identical.
\begin{equation}
\frac{d\sigma (pp \to pp J/\Psi, W)}{dy} =
\frac{d\sigma (pp \to p \bar p J/\Psi)}{dy}
\label{dsig_dy_pp_vs_ppbar}
\end{equation}
It is interesting to stress in this context that it is not the case for
transverse momentum distribution of $J/\psi$, where
\begin{equation}
\frac{d\sigma(pp \to pp J/\Psi,W)}{d^2\bp_{V}}  \ne
\frac{d\sigma (p\bar p \to p \bar p J/\Psi,W)}{d^2\bp_{V}} \; .
\label{dsig_dy_pp_vs_ppbar}
\end{equation}
This is demonstrated in Fig.\ref{fig:dsig_dpvt2}, where we see,
that at small transverse momenta of the vector meson, the interference
enhances the cross section in $pp$ collisions and depletes it
in $p\bar p$ collisions. It is a distinctive feature of the 
mechanism discussed here, that vector mesons are produced with
very small transverse momenta.
The difference between proton-antiproton and proton-proton
collisions survives even at large rapidities of $J/\psi$.
When integrated over the $J/\psi$ transverse momentum, and in
absence of absorptive corrections, cross sections
will again be identical in the $pp$ and $p \bar p$ cases.

\subsection{Distributions of (anti--)protons}

\begin{figure}[!h]  
 \centerline{\includegraphics[width=0.4\textwidth]{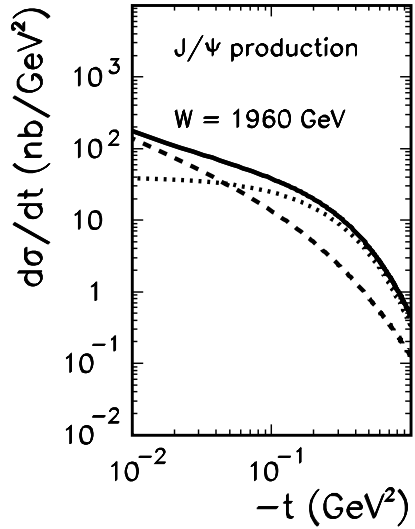}}
   \caption{ \label{fig:dsig_dt12}
\small $d \sigma / dt_{1/2}$ as a function of Feynman $t_{1/2}$
for W = 1960 GeV. The photon-exchange (dashed) and pomeron-exhange
(dotted) contributions are shown in addition. 
No absorption corrections were included.}
\end{figure}

Now we shall proceed to distributions related to (anti--)protons.
In Fig.\ref{fig:dsig_dt12} we show distributions in the transferred
momenta squared (identical for $t_1$ and $t_2$).
We show separately the contributions of $\gamma\Pom$ and $\Pom\gamma$
exchanges. The figures clearly display the strong photon--pole enhancement
at very small $t$.


\begin{figure}[!h]   
\includegraphics[width=0.45\textwidth]{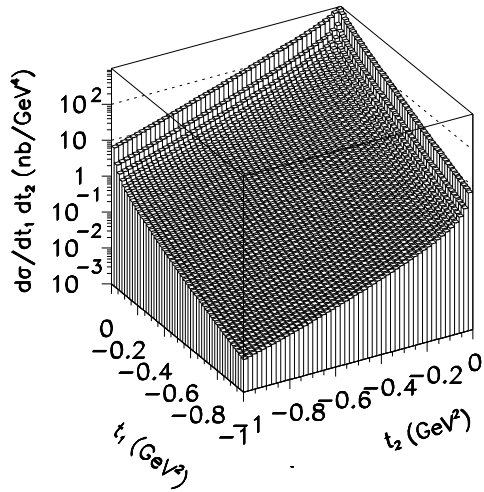}
\caption{ Two-dimensional distribution in $t_1$ and $t_2$ for
the Tevatron energy W = 1960 GeV. In this calculation a full range
of the $J/\psi$ rapidities was included.
No absorption corrections were included here.}
\label{fig:map_t1t2}
\end{figure}

In order to better understand the distributions in $t_1$ or $t_2$
in Fig.\ref{fig:map_t1t2} we show how $t_1$ and $t_2$ are correlated.
Here we do not make any restrictions on the rapidity range.
The significant enhancements of the cross section in the form
of ridges along $t_1 \sim$ 0 and $t_2 \sim$ 0 are again due
to the massless photon--exchange, and most of the integrated cross section
comes from these regions. The pomeron-odderon and odderon-pomeron exchange
contributions considered in Ref.\cite{BMSC07} would not exhibit such
a significant local enhancements and would be smeared over broader range
in the $(t_1,t_2)$ space. Therefore in the dedicated searches for
the odderon exchange upper cuts on $t_1$ and $t_2$ should be imposed,
and $t_{upper}= -0.2$ GeV seems to be a good choice.


\begin{figure}[!h]   
 \centerline{\includegraphics[width=0.5\textwidth]{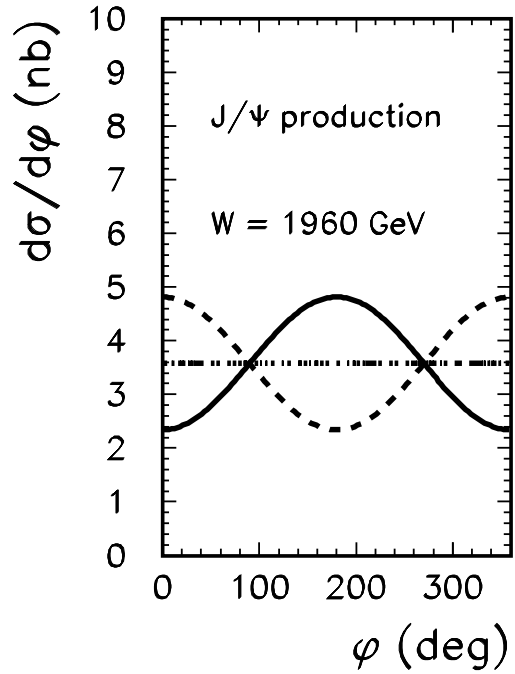}}
   \caption{ \label{fig:dsig_dphi_w1960}
\small  $d \sigma / d\phi$ as a function of $\phi$ for W = 1960 GeV.
The solid line corresponds to a coherent sum of amplitudes
whereas the dashed line to incoherent sum of both processes.
No absorption corrections were included here.}
\end{figure}

We repeat that the reaction considered in this paper
leads to azimuthal correlations between outgoing proton
and antiproton. In Fig.\ref{fig:dsig_dphi_w1960} we show the corresponding
angular distribution for proton-antiproton collision (solid line).
For reference we also show, by the dotted line, the incoherent sum of
the $\gamma \Pom$ and $\Pom \gamma$ mechanisms. 
The distribution for proton-proton collisions (dashed line) is shown for comparison
alsoat the Tevatron energy.
Clearly the interference terms in both reactions are in opposite phase due
to different electric charges of proton and antiproton.
In the absence of absorptive corrections, we have
\begin{equation}
\frac{d \sigma}{d \phi} = A \pm B \cos\phi
\end{equation}
for $p p$ (+) and $p \bar p$ (-) collisions respectively.
The interference effect (B/A) here is at the level of $\sim$ 40--50 \%.


\begin{figure}[!h]   
\includegraphics[width=0.45\textwidth]{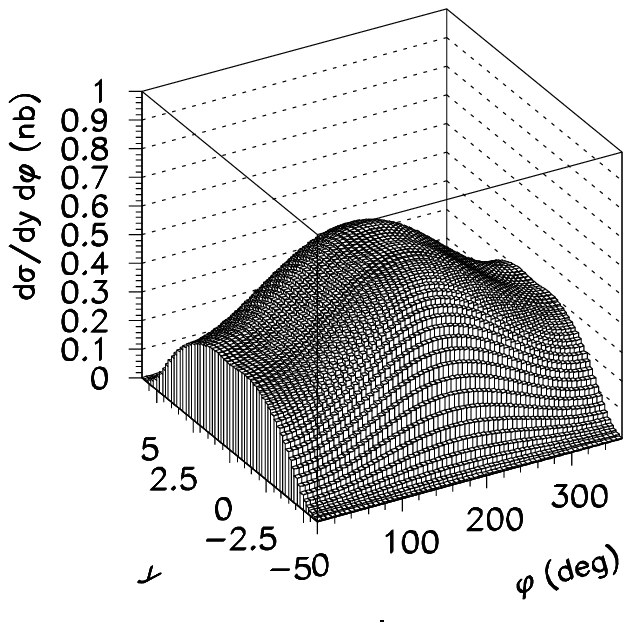}
\includegraphics[width=0.45\textwidth]{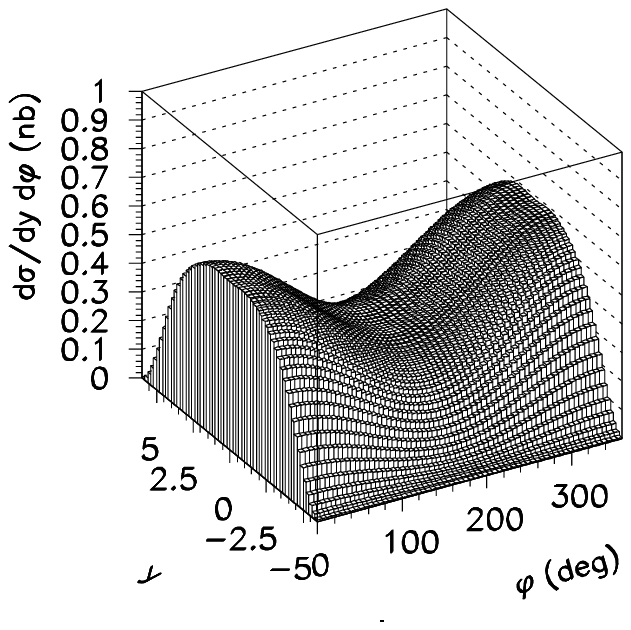} 
   \caption{ \label{fig:map_yphi.eps}
\small $d \sigma / dy d \phi$ for W = 1960 GeV and for $p \bar p$ (left panel) 
and $p p$ (right panel) collisions.
No absorption corrections were included here.}
\end{figure}

In Fig.\ref{fig:map_yphi.eps} we show the two-dimensional distributions
differentially in both rapidity and azimuthal angle. 
Interestingly, the interference effect is significant over broad range
of $J/\psi$ rapidity, which is reflected in the fact that even at large
$J/\psi$ rapidities one observes ansisotropic distributions in
the azimuthal angle. 


\begin{figure}[!h]   
\includegraphics[width=0.5\textwidth]{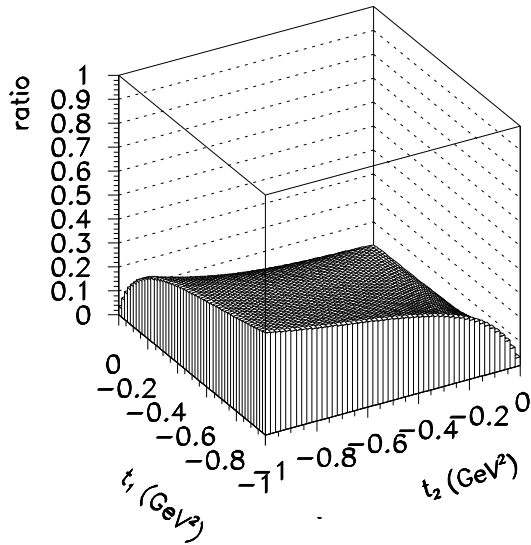}
   \caption{ \label{fig:sftosp}
\small
The ratio of helicity-flip to helicity-preserving contribution
as a function of $t_1$ and $t_2$.}
\end{figure}

Up to now we have considered only spin-preserving contributions.
Now we wish to show the effect of electromagnetic spin-flip discussed in
the previous section.
In Fig.\ref{fig:sftosp} we show the ratio of helicity-flip to
helicity-preserving contribution. The ratio is a rather flat function
of $t_1$ and $t_2$. At $t_1 = -1$ GeV$^2$ and $t_2 = -1$
GeV$^2$ the ratio reaches about 0.4.

\subsection{Absorption effects}

Now we will show the effect of absorptive corrections
discussed in section \ref{Absorption} on various differential distributions.

\begin{figure}[!h]   
\includegraphics[width=0.45\textwidth]{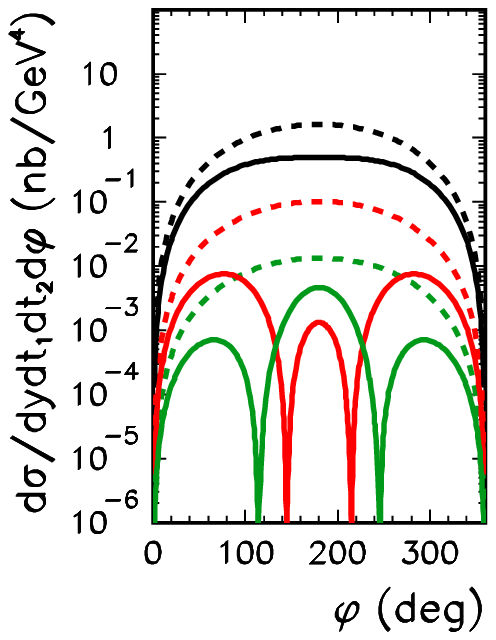}
\includegraphics[width=0.45\textwidth]{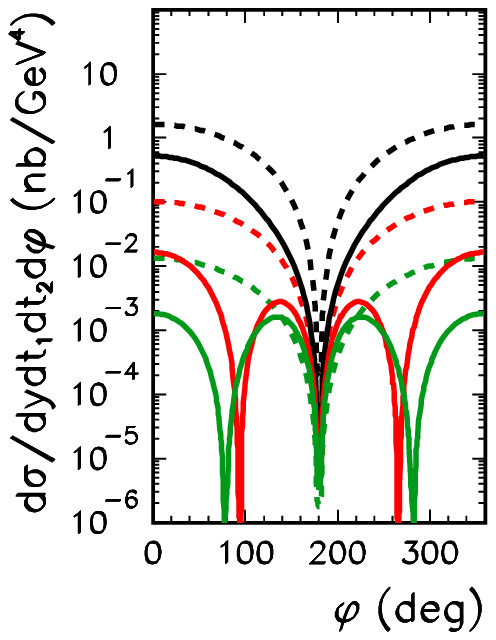}
   \caption{ \label{fig:M2_phi_w1960}
Fully differential cross section $d \sigma/dy dt_1 dt_2 d \phi$
as a function of $\phi$ for $y$=0 and different
combinations of $t_1 = t_2 \,  (-0.1,-0.3,-0.5$ GeV$^2$ (from top to bottom))
for $p \bar p$ (left panel) and $p p$ (right panel) reactions.
The solid lines include rescattering, while the dashed lines
correspond the Born--level mechanism only.
}
\end{figure}
\begin{figure}[!h]   
\includegraphics[width=0.4\textwidth]{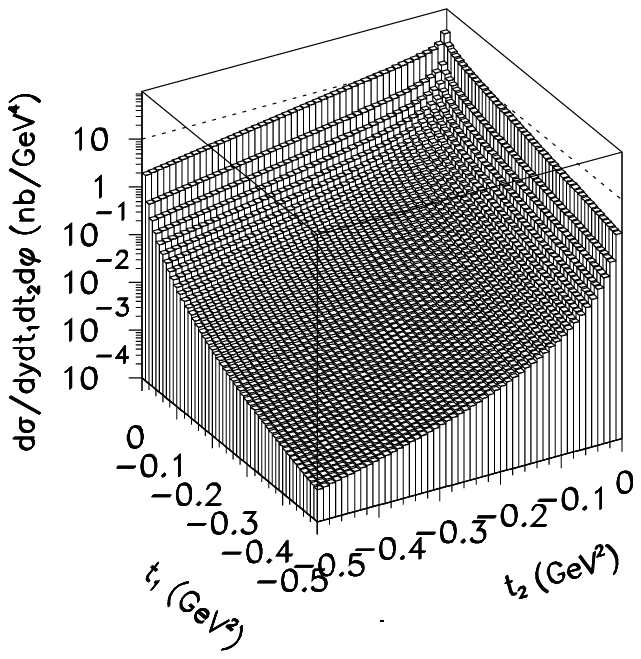}
\includegraphics[width=0.4\textwidth]{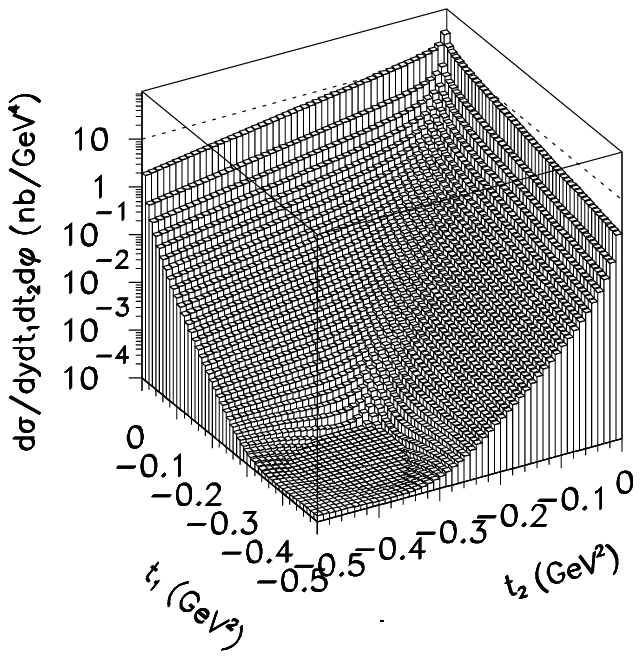}
   \caption{ \label{fig:M2_t1t2_w1960}
The fully differential cross section
$d \sigma/dy dt_1 dt_2 d \phi$
as a function of $t_1$ and $t_2$ for $y$=0 and $\phi = \pi/2$
for $p \bar p$ (left panel) and $p p$ (right panel) reactions.
}
\end{figure}
\begin{figure}[!h]   
\includegraphics[width = \textwidth]{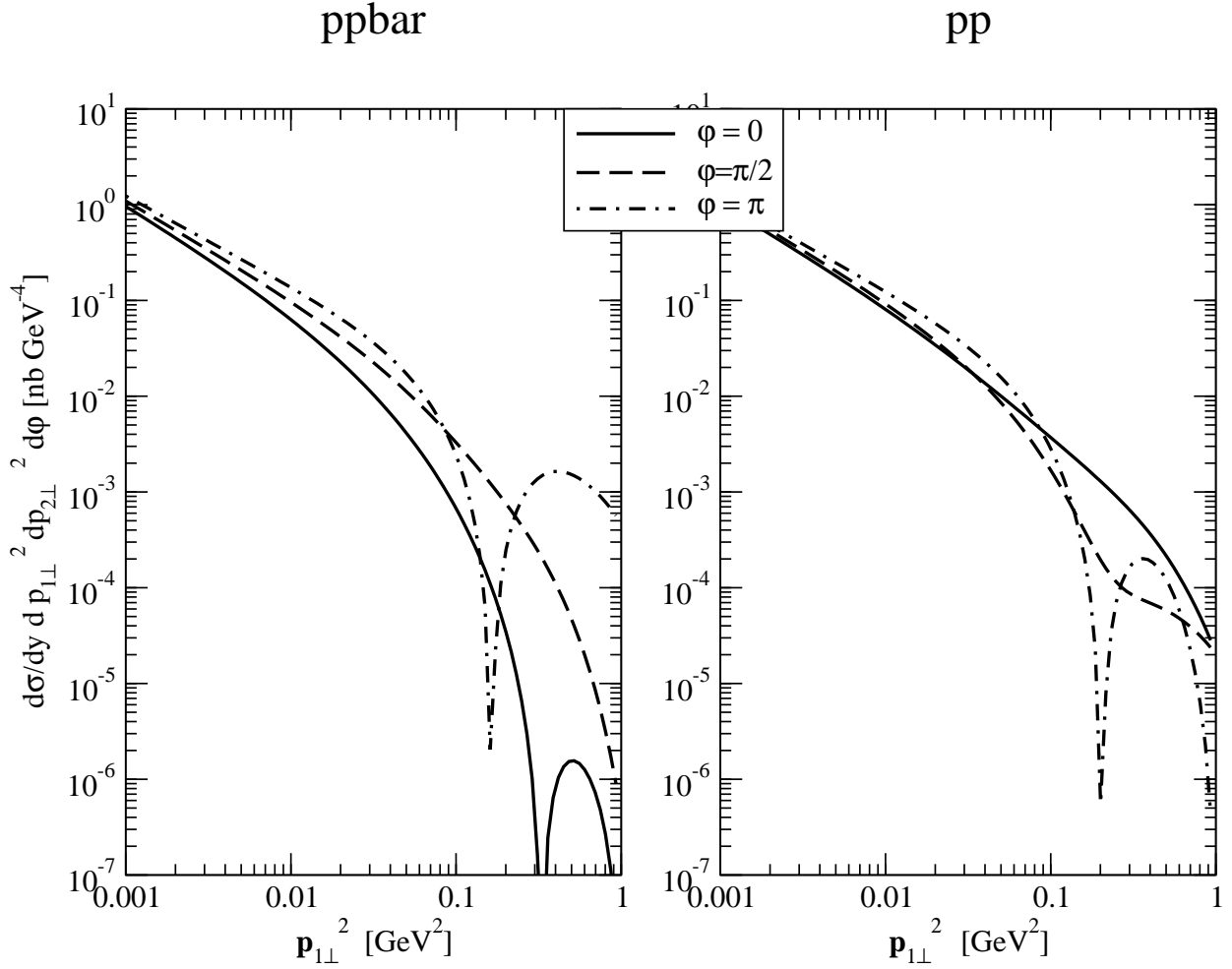}
   \caption{ \label{fig:dsig_scan}
Fully differential cross section $d \sigma/dy d\bp^2_1 d\bp^2_2\ 
d \phi$ as a function of $\bp^2_1$ at $y=0$ and 
$\bp^2_2 = 1$GeV$^2$ for $p \bar p$ (left panel) and 
$pp$ (right panel) collisions at $W = 1960$ GeV. 
Absorptive corrections (elastic rescattering) are included.
Solid, dashed, and dash--dotted lines refer to different values of
the azimuthal angle $\phi$ between outgoing (anti-)protons.
}
\end{figure}
\begin{figure}[!h]   
\includegraphics[width= \textwidth]{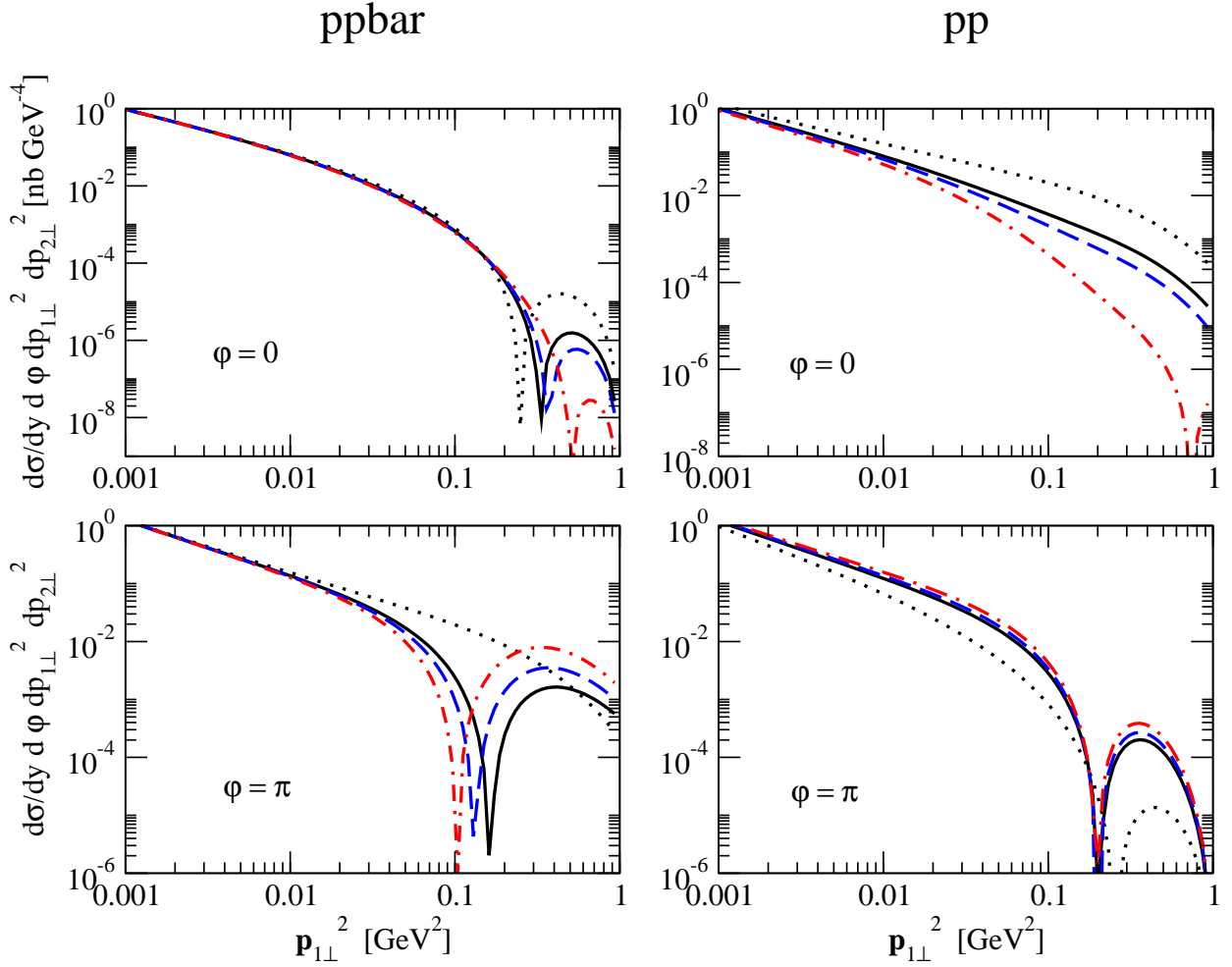}
 \caption{ \label{fig:dsig_scan_absorption}
Fully differential cross section $d \sigma/dy d\bp^2_1 d\bp^2_2 
d \phi$ as a function of $\bp^2_1$ at $y=0$ and $\bp^2_2=1$GeV$^2$
for $p \bar p$ (left panel) and $pp$ (right panel) collisions
at $W = 1960$ GeV. The azimuthal angle $\phi$ is taken $\phi = 0$
in the top row and $\phi = \pi$ in the second row.
Shown by the dotted curve is the Born-level cross section, without 
absorptive corrections included. The solid curve shows the result
with elastic rescattering between (anti-) protons included, while
rescattering has been enhanced by a factor $\lambda = 1.2$ in the 
amplitude for the dashed curve, and by a factor $\lambda = 1.5$
for the dash--dotted curve.
}
\end{figure}
Let us start from the presentation of the effects of absorption 
for selected points in phase space.
In Fig.\ref{fig:M2_phi_w1960} we show 
the fully differential cross section $d \sigma/dy dt_1 dt_2 d \phi$
as a function of $\phi$ for selected (fixed) values of $t_1$, $t_2$
and for $y$ = 0. We show results for $p \bar p$ (left panel)
and $pp$ (right panel) collisions for the same center-of-mass
energy $W$ = 1960 GeV. While at smaller $t_{1,2}$ we observe a smooth 
reduction of the Born--level result, absorptive 
corrections induce a strong $\phi$--dependence at larger $t_{1,2}$.
The positions of the diffractive minima which appear as a consequence
of cancellations of the Born and rescattering amplitudes
move with the value of $t \equiv t_1 = t_2$.
In Fig.\ref{fig:M2_t1t2_w1960} we present fully differential cross
section as a function
of $t_1$ and $t_2$ for $y$=0 and $\phi= \pi/2$. 
For proton-proton scattering we observe clearly
a diffractive minimum for $t_1 = t_2 \approx$ 0.4 GeV$^2$.
In Fig. \ref{fig:dsig_scan} we show the fully differential cross section
as a function of the transverse momentum squared $\bp_1^2$ at a fixed
value $\bp_2^2 = 1 $GeV$^2$ at rapidity $y = 0$. The rich structure as
a function of transverse momenta and azimuthal angles is also revealed
by this plot. The plots in Fig.\ref{fig:dsig_scan_absorption} 
give an idea, to which extent the diffractive dip--bump structure 
depends on the details of our treatment of absorption.
Here we show, by the dotted line, the cross section calculated
for the Born--level amplitude. The solid line shows the result
with elastic scattering included, and for the dashed and dash--dotted 
lines we enhanced the rescattering amplitude $T$ of section \ref{Absorption}
by a factor $\lambda = 1.2$ and $\lambda = 1.5$ respectively.
The region of very small $\bp_1^2$ is entirely
insensitive to rescattering, reflecting the ultraperipheral nature
of photon exchange. The diffractive dip--bump structure, situated
at larger transverse momenta reveals a dependence on the strength
of absorptive corrections. This concerns the position of dips as well
as the strength of the cross section in various windows of
phase space. 

\begin{figure}[!h]   
\includegraphics[width=0.4\textwidth]{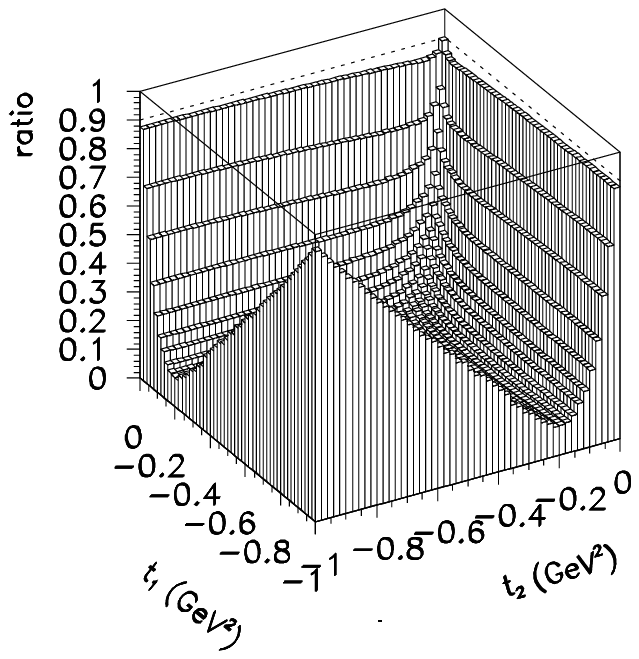}
\includegraphics[width=0.4\textwidth]{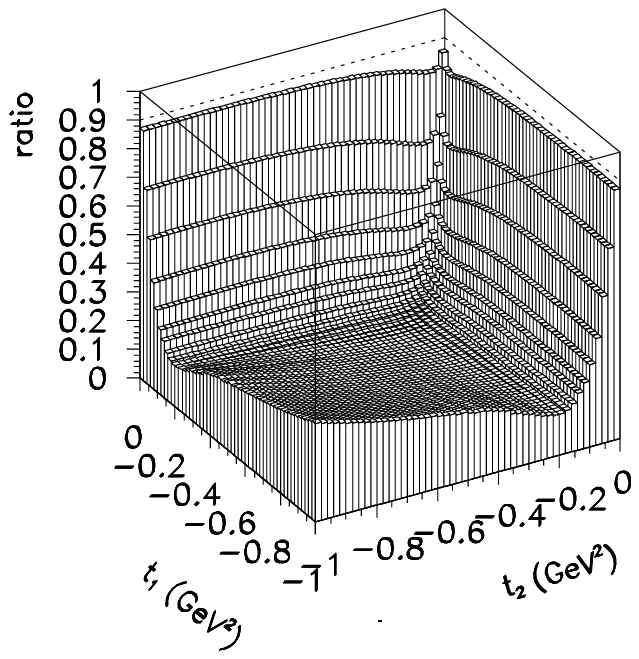}
   \caption{ \label{fig:ratio_absorption_t1t2}
The ratio of the cross sections with absorption to that without
absorption for $p \bar p$ (left panel) and $p p$ (right panel) scattering.
Here the integration over $y$ and $\phi$ was performed.
}
\end{figure}

\begin{figure}[!h]   
\includegraphics[width=0.4\textwidth]{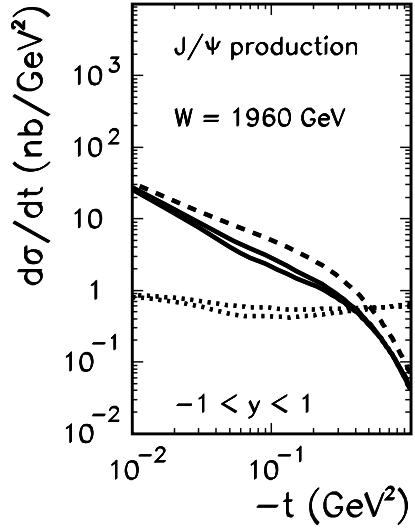}
   \caption{ \label{fig:dsig_dt_absorption}
Differential cross section
$d \sigma/dt$, integrated over $-1<y<1$ for $p\bar p$ collisions.
Dashed line: no absorptive corrections. Solid lines show the cross 
section with absorptive corrections included. Upper solid line: purely
elastic rescattering. Lower solid line: rescattering enhanced by a factor 
$\lambda = 1.5$ in the amplitude. The dotted lines show the respective
ratios $d \sigma^{Born + Rescatt.}/ d\sigma^{Born}$.
}
\end{figure}

\begin{figure}[!h]   
\includegraphics[width=\textwidth]{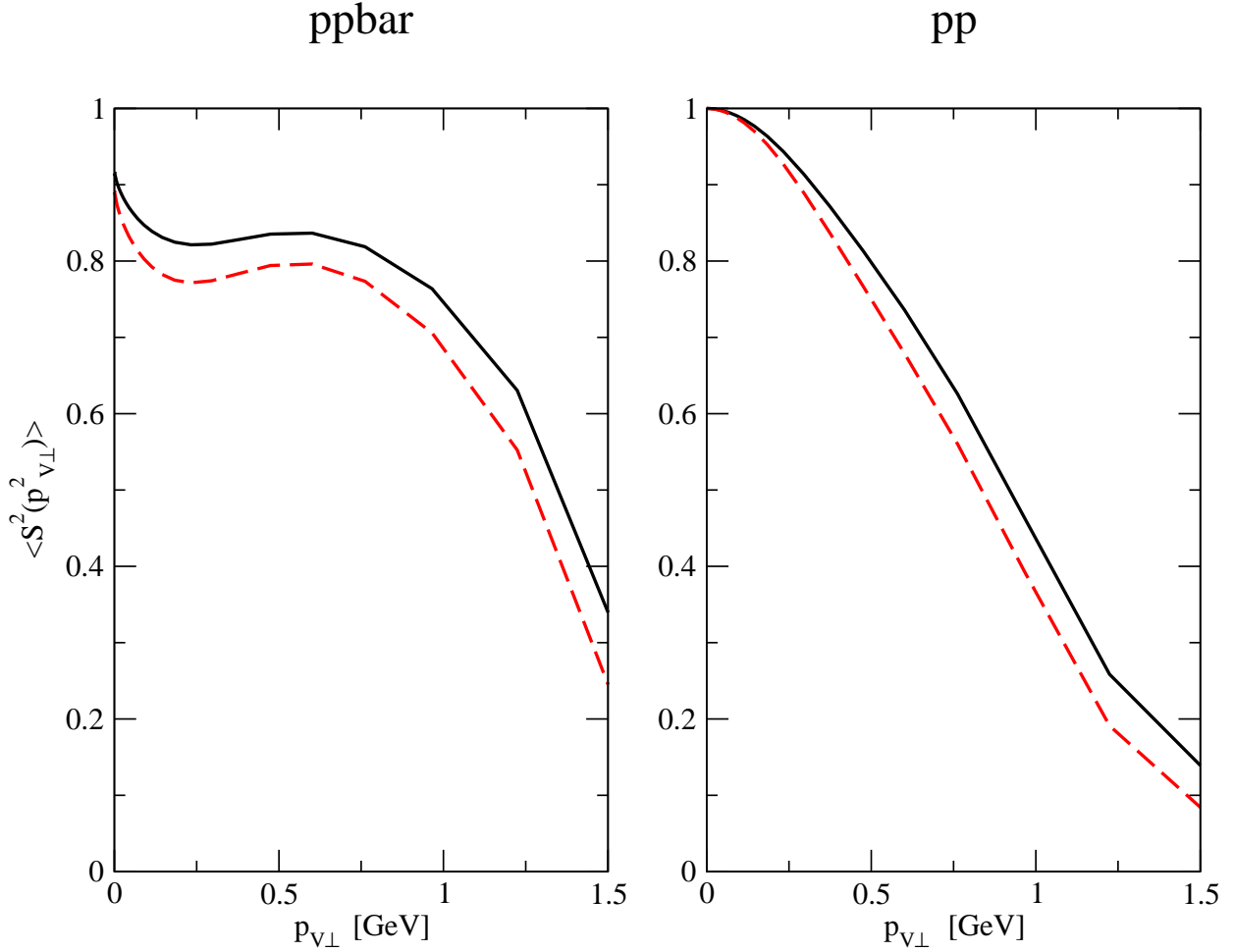}
   \caption{ \label{fig:ratio_pv_abs}
The suppression factor $<S^2(\bp_V^2)>$ at $y=0$ as a function of the 
$J/\Psi$ transverse momentum, for $p\bar p$ (left panel) and
$pp$ (right panel) collisions at $W = 1960$ GeV.
}
\end{figure}

The sensitivity to rescattering is however washed
out in integrated observables -- clearly the contribution
from low transverse momenta, which is not strongly affected by
rescattering, is large. This becomes apparent in 
Figs. \ref{fig:ratio_absorption_t1t2}, \ref{fig:dsig_dt_absorption} 
and \ref{fig:ratio_pv_abs}.

In Fig.\ref{fig:ratio_absorption_t1t2} we show
the ratio of the cross section with absorption to that without
absorption as a function of $t_1$ and $t_2$, again for $p \bar p$ (left)
and $p p$ (right). As a consequence of averaging over different 
phase-space configurations the diffractive minima disappeared.
Again the plot reflects, that due to the ultraperipheral nature of 
the photon--exchange mechanism, absorption is negligible at very small 
$t_{1,2}$, and rises with $t_1$ and/or $t_2$. 
On average, absorption for the $p \bar p$ reaction is
smaller than for the $p p$ case.

It is important to stress again, that the absorptive corrections
in differential cross sections cannot be accounted for by simply
a constant suppression factor, but show a lively dependence over
phase space. In Fig. \ref{fig:dsig_dt_absorption} we show the differential 
cross section $d\sigma/dt$. The dashed line shows the result without 
absorption, the solid lines include absorptive corrections.
They differ by a factor $1.5$, by which rescattering had been enhanced
in the lower curve. This enhancement of rescattering shows a modest effect,
quite in agreement with the expectation mentioned above.
The dependence of absorption on $t$ is quantified by the ratio 
of full and Born--level cross sections shown by the dotted lines.

In Fig. \ref{fig:ratio_pv_abs} we show the suppression factor
\begin{equation}
<S^2(\bp_V^2)> = {d\sigma^{{\mathrm{Born + Rescatt.}}}/d\bp_V^2dy 
\over d\sigma^{{\mathrm{Born}}}/d\bp_V^2dy} \, ,
\end{equation}
as a function of the $J/\Psi$ transverse momentum at $y=0$.
It is important to emphasize once more the strong functional
dependence on $\bp_V$, which is different for $p\bar p$ and
$pp$ collisions. Again, playing with the strength of absorptions
shows a modest effect.

The different behaviour of absorptive corrections in $pp$ and
$p\bar p$ collisions is an interesting observation. It derives
from the fact that rescattering corrections lift the cancellation
of the interference term after azimuthal integration.
Finally, let us comment on the expected reduction of rapidity 
distributions from absorptive corrections. These are, finally,
rather flat functions of $y$. For the ratio
\begin{equation}
<S^2(y)> = {d\sigma^{{\mathrm{Born + Rescatt.}}}/dy \over d\sigma^{{\mathrm{Born}}}/dy}
\, , 
\end{equation}
we obtain, in $p\bar p$ collisions
\begin{equation}
<S^2(y = 0)>\Big|_{p\bar p} \approx 0.9 \, , \, 
<S^2(y = 3)>\Big|_{p\bar p} \approx 0.8 \, , \, 
\end{equation}
and for $pp$ collisions
\begin{equation}
<S^2(y = 0)>\Big|_{p p} \approx 0.85 \, , \, 
<S^2(y = 3)>\Big|_{p p} \approx 0.75 \, . \, 
\end{equation}
We note that this results in a small charge asymmetry
\begin{equation}
 {
d\sigma(p \bar p) / dy - d\sigma (pp) / dy 
\over 
d\sigma(p \bar p) / dy + d\sigma (pp) / dy} \approx  2 \div 3 \,  \% \, ,
\end{equation}
which derives entirely from absorptive corrections.

\section{Conclusions}
In this paper we have calculated differential cross sections for
exclusive $J/\psi$ production via photon-pomeron($\gamma \Pom$) 
and pomeron-photon ($\Pom\gamma$)
exchanges at RHIC, Tevatron and LHC energies. 
Measurable cross sections were obtained in all cases.
We have obtained an interesting azimuthal-angle correlation pattern
due to the interference of the $\gamma\Pom$ and $\Pom\gamma$ 
mechanisms. The interference effect survives over almost the 
whole range of $J/\psi$ rapidities.
At the Tevatron energy one can potentially study the exclusive
production of $J/\psi$ at the photon-proton center-of-mass energies
70 GeV $ < W_{\gamma p} < $ 1500 GeV, i.e. in the unmeasured region
of energies, much larger than at HERA. At LHC this would be correspondingly
200 GeV $ < W_{\gamma p} < $ 8000 GeV. At very forward
rapidities this is an order of magnitude more than possible
with presently available machines.
Due to the photon--pole, the differential cross section 
is concentrated in the region of very small $t_1$ or/and $t_2$. 
Imposing cuts on $t_1$ and $t_2$ lowers the cross section considerably.
Electromagnetic helicity-flip processes play some role
only when both $|t_1|$ and $|t_2|$ are large, that is in a region where
also the hypothetical hadronic, Odderon exchange, contribution can be 
present.
It is a distinctive feature of the production mechanism, that
mesons are produced at very small transverse momenta, where 
the interference of $\gamma\Pom$ and $\Pom \gamma$ mechanisms
induces a strongly different shape of vector--meson $\bp_V^2$--
distributions in $pp$ vs. $p\bar p$ collisions.
We also estimated absorption effects on various distributions. 
In some selected configurations the absorptive corrections
lead to the occurence of diffractive minima.
Naturally, the exact place of diffractive minima depends on the values
of the model parameters, but they are washed out when integrated
over the phase space or even its part.
Absorptive corrections for differential distributions are lively functions
of transverse momenta, and cannot be accounted for simply by constant
suppression factors.  
We have found that absorptive corrections induce a small charge--asymmetry
in rapidity distributions and total production cross sections.
In the present paper we have concentrated on exclusive production of
$J/\psi$ at energies $\sqrt{s} > 200$ GeV.
The formalism used here can be equally well applied to exclusive
production of other vector mesons, such as $\phi$, $\Upsilon$
as well as to the lower energies of
e.g. FAIR, J-PARC, RHIC. For $J/\psi$--production, especially recent 
parametriztions \cite{FJMPP07}
of the photoproduction cross section from threshold to the highest
energies may prove useful.
We leave such detailed analyses for separate studies.
The processes considered here are also interesting in the context
of recently proposed searches for identifying the Odderon. 
We find that the region of midrapidities ($ -1  < y <  1 $) and
$t_1, t_2 < -0.2$ GeV$^2$ seems the best in searches for the odderon exchange.
Should data reveal deviations from the conservative predictions given by us,
a detailed differential analysis of both, photon and odderon--echange processes, including their interference, in $y, t_1, t_2, \phi$ will be called upon.

\section{Acknowledgments}
We thank Mike Albrow for useful comments regarding the experimental
possibilities of exclusive $J/\Psi$--production at the Tevatron.
We are indebted to Tomasz Pietrycki for help in preparing some figures.
This work was partially supported by the grant
of the Polish Ministry of Scientific Research and Information Technology
number 1 P03B 028 28.

\section{Appendix}

In Ref.\cite{H1_JPsi} the differential cross section $\frac{d\sigma}{dt}$
for the reaction $\gamma^* p \to J/\psi p$ was parametrized as
\begin{equation}
\frac{d\sigma}{dt}(W,t,Q^2) = \frac{d\sigma}{dt}\Big|_{t=0,W=W_0}
\left( \frac{W}{W_0} \right)^{4 (\alpha(t)-1)} \exp( B_0 t )
\left( \frac{m_{J/\psi}^2}{m_{J/\psi}^2+Q^2} \right)^n \; ,
\label{H1_cross_section}
\end{equation}
where $\alpha(t) = \alpha_0 + \alpha' t$ .
The values of parameters found from the fit to the data are:
$\frac{d\sigma}{dt}|_{t=0,W=W_0}$ = 326 nb/GeV$^2$, $W_0$ = 95 GeV,
$B_0$ = 4.63 GeV$^{-2}$, $\alpha_0$ = 1.224, $\alpha'$ = 0.164 GeV$^{-2}$
, $n$ = 2.486.

Assuming the dominance of the helicity-conserving transitions, and
neglecting the real part, one can
write
\begin{equation}
{\cal M}(s,t,Q^2) =
\delta_{\lambda_{\gamma} \lambda_V} 
\delta_{\lambda_{p} \lambda_{p'}}
 i s \sqrt{ 16 \pi \frac{d\sigma}{dt}\Big|_{t=0,W=W_0}  }
\left( \frac{s}{W_0^2} \right)^{\alpha(t)-1} \exp( B_0 t / 2 )
\left( \frac{m_{J/\psi}^2}{m_{J/\psi}^2+Q^2} \right)^{n/2} \; ,
\label{H1_amplitude}
\end{equation}
identical for each combination of particle helicities.
In our case of hadroproduction the amplitude is a function of
either $(s_1,t_1,Q_2^2)$ or $(s_2,t_2,Q_1^2)$.


\end{document}